\documentclass[onecolumn,tightenlines,showpacs,nofootinbib,preprintnumbers,
amsmath,amssymb]{revtex4}

\usepackage[usenames,dvipsnames]{color}
\usepackage{caption2}
\usepackage{dcolumn}
\usepackage{graphicx}
\usepackage{bm}
\usepackage{lscape}

\usepackage{setspace}
\begin{document}
\begin{spacing}{1.5}

\title{Effect of resonance for $CP$ asymmetry of the decay process $\bar{B}_{s}\rightarrow P\pi^+\pi^-$ in perturbative QCD}

\author{Gang L\"{u}$^{1}$\footnote{Email: ganglv66@sina.com}, Yu-Ting Wang$^{1}$$^{2}$\footnote{Email: 1206166292@qq.com},Qin-Qin Zhi$^{1}$\footnote{Email: zhiqinqin11@163.com}}

\affiliation{\small $^{1}$College of Science, Henan University of Technology, Zhengzhou 450001, China\\
\small $^{2}$ Institute of High Energy Physics Chinese Academy of Sciences, Beijing 100049, China\\
}

\begin{abstract}
In the framework of Perturbative QCD (PQCD) approach we study the direct $CP$ asymmetry for the decay channel
$\bar{B}_{s}\rightarrow P\pi^+\pi^-$ around the resonance range via the $\rho-\omega$ mixing mechanism (where P refer to pseudoscalar meson). We find that the $CP$ asymmetry can be
enhanced by $\rho-\omega$ mixing when the masses of the $\pi^+\pi^-$ pairs are at the area of $\rho-\omega$ resonance, and the maximum $CP$ asymmetry can
reach 59{\%} for the relevant decay channels.
\end{abstract}


\maketitle

\section{\label{intro}Introduction}
The rich data from $B$ meson factories make the study of $B$ physics a very hot topic. A lot of research has been made, especially for $CP$ asymmetry.
$CP$ asymmetry is an important area in test of the Standard Model (SM) and searching new physics signals.
The detection of  Cabibbo-Kobayashi-Maskawa (CKM) matrix elements play an important role in understanding
of $CP$ asymmetry. The nonleptonic decay of $B$ meson is expected to be ideal decay process in searching $CP$ asymmetry.
Direct $CP$ asymmetry in $B$ meson decay channel arises from weak phase and strong phase differences.
In SM, the weak phase is responsible for the $CP$ asymmetry by CKM matrix \cite{cab,kob}.
Meanwhile, the large strong phase is needed for producing $CP$ asymmetry which comes from QCD correction.
Recently, the large $CP$ asymmetry was found by the
LHCb Collaboration in the three-body
decay channels of $B^{\pm}\rightarrow \pi^{\pm}\pi^{+}\pi^{-}$
and $B^{\pm}\rightarrow K^{\pm}\pi^{+}\pi^{-}$\cite{J-R.A}.
Hence, more attention about $CP$ asymmetry has been focused on the three body decay channels of $B$ meson.

Direct $CP$ asymmetry arises from the weak phase difference and the strong phase difference.
The weak phase difference is determined by the CKM matrix elements, while the
strong phase can be produced by the hadronic matrix and
interference between intermediate states. The vacuum polarisation of photon are described  by coupling the vector meson
in the vector meson dominance (VMD) model. The strength of coupling of the $\omega$ meson to the photon is weak comparing with the $\rho$ meson \cite{HB1997}.
However, the strong interaction enhances the $\pi^{+}\pi^{-}$ pair production amplitudes in the $\rho$ and $\omega$ resonance region.
$\rho-\omega$ interference presents the large contribution for the process of $e^{+}e^{-}\rightarrow \pi^{+}\pi^{-}$ due to
the isospin-breaking effects. Since the strong phase exist, the $\rho$ and $\omega$ interference can affect the direct CP asymmetry and present
the sizeable contribution.

The direct $CP$ asymmetry is discussed via $\rho-\omega$ interference in $B$ decays by the the naive factorization approach \cite{Ryoji1996}.
But the method bases on the assumption of no strong rescattering, and can not predict direct CP asymmetry effectively.
Recently, the CP asymmetry of charmless three-body B-decay is presented in the leading term of QCD factorization by model dependent approach,
where focus on the local $CP$ asymmetry \cite{Rebecca2017}. The direct $CP$ asymmetry of the quasi-two-body decay of $B\rightarrow P\rho\rightarrow P\pi\pi$
is calculated in perturbative QCD approach, where does not taking into account the resonance effects \cite{Yali2017}.
In our opinion, $B \rightarrow P\pi\pi$ have effectively three contributions around the $\rho$ resonance:
(a) $B\rightarrow P\rho\rightarrow P \pi \pi$, (b)$B\rightarrow P \omega \rightarrow P\rho\rightarrow P \pi \pi$, and
(c)$B \rightarrow P \omega \rightarrow P \pi \pi$. Roughly speaking, the amplitudes of their contributions: $a>b>c$. We have
absorbed (c) into (b) effectively, which is just the (effective) $\rho-\omega$ mixing parameter:$\tilde{\Pi}_{\rho\omega}$.

The hadronic matrix elements can be calculated by the factorization approach introducing the strong phase.
Adding the QCD corrections, the different dynamic methods are given
based on the leading power of $1/m_b$ ($m_b$ is b quark mass).
The non-leptonic weak decay amplitudes of B mesons can be calculated by the perturbative QCD (PQCD) approach
taking into account transverse momenta \cite{AC,AG,YA,LKM}.
In the PQCD approach, the hard interaction consisting of
six quark operator dominants the decay amplitude from short distance.
The nonperturbative dynamics are included in the meson
wave function which can be extracted from experiment.
Finally, we obtain new large strong phases by the phenomenological
mechanism of $\rho-\omega$ mixing and the dynamics of the PQCD approach.
The large $CP$ asymmetry may be obtained by the resonant region due to the strong phase.

The remainder of this paper is organized as follows. In Sec.
\ref{sec:hamckm} we present the form of the effective Hamiltonian.
In Sec. \ref{sec:cpv1} we give the calculating formalism of $CP$ asymmetry from $\rho-\omega$ mixing
in $\bar{B}_{s}\rightarrow P\pi^+\pi^-$.
Input parameters are presented in Sec.\ref{input}.
We present the numerical results in Sec.\ref{sec:numerical}.
Summary and discussion are included in
Sec. \ref{sec:conclusion}. The related function defined in the text are given
in the Appendix.

\section{\label{sec:hamckm}The effective hamiltonian}
Based on the expansion of the operator product, the effective weak Hamiltonian can be written as \cite{GW}
\begin{eqnarray}
{\cal H}_{\Delta B=1} = {G_F\over \sqrt{2}}[
V_{ub}V^*_{ud}(c_1 O^u_1 + c_2 O^u_2)    \nonumber   \\
 - V_{tb}V^*_{td}\sum^{10}_{i=3} c_i O_i] + H.C.,\;
\label{2a}
\vspace{2mm}
\end{eqnarray}
where $G_F$ represents Fermi constant, $c_i$ (i=1,...,10) are the Wilson coefficients, $V_{ub}$,
$V_{ud}$, $V_{tb}$ and $V_{td}$ are the CKM matrix elements. The
operators $O_i$ have the following forms:
\begin{eqnarray}
O^{u}_1&=& \bar d_\alpha \gamma_\mu(1-\gamma_5)u_\beta\bar
u_\beta\gamma^\mu(1-\gamma_5)b_\alpha,\nonumber\\
O^{u}_2&=& \bar d \gamma_\mu(1-\gamma_5)u\bar
u\gamma^\mu(1-\gamma_5)b, \nonumber
\label{2b1}
\vspace{2mm}
\end{eqnarray}
\begin{eqnarray}
O_3&=& \bar d \gamma_\mu(1-\gamma_5)b \sum_{q'}
\bar q' \gamma^\mu(1-\gamma_5) q',\nonumber\\
O_4 &=& \bar d_\alpha \gamma_\mu(1-\gamma_5)b_\beta \sum_{q'}
\bar q'_\beta \gamma^\mu(1-\gamma_5) q'_\alpha,\nonumber\\
O_5&=&\bar d \gamma_\mu(1-\gamma_5)b \sum_{q'} \bar q'
\gamma^\mu(1+\gamma_5)q',\nonumber\\
O_6& = &\bar d_\alpha \gamma_\mu(1-\gamma_5)b_\beta \sum_{q'}
\bar q'_\beta \gamma^\mu(1+\gamma_5) q'_\alpha,\nonumber\\
O_7&=& \frac{3}{2}\bar d \gamma_\mu(1-\gamma_5)b \sum_{q'}
e_{q'}\bar q' \gamma^\mu(1+\gamma_5) q',\nonumber\\
O_8 &=&\frac{3}{2} \bar d_\alpha \gamma_\mu(1-\gamma_5)b_\beta \sum_{q'}
e_{q'}\bar q'_\beta \gamma^\mu(1+\gamma_5) q'_\alpha,\nonumber\\
O_9&=&\frac{3}{2}\bar d \gamma_\mu(1-\gamma_5)b \sum_{q'} e_{q'}\bar q'
\gamma^\mu(1-\gamma_5)q',\nonumber\\
O_{10}& = &\frac{3}{2}\bar d_\alpha \gamma_\mu(1-\gamma_5)b_\beta \sum_{q'}
e_{q'}\bar q'_\beta \gamma^\mu(1-\gamma_5) q'_\alpha,\nonumber\\
&&
\label{2b}
\vspace{2mm}
\end{eqnarray}
where $\alpha$ and $\beta$ are color indices, and $q^\prime=u, d$
or $s$ quarks. In Eq.(\ref{2b}) $O_1^u$ and $O_2^u$ are tree
operators, $O_3$--$O_6$ are QCD penguin operators and $O_7$--$O_{10}$ are
the operators associated with electroweak penguin diagrams.

we can obtain numerical values of $c_i$. When $c_i(m_b)$ \cite{LKM},
\begin{eqnarray}
c_1 &=&-0.2703, \;\; \;c_2=1.1188,\nonumber\\
c_3 &=& 0.0126,\;\;\;c_4 = -0.0270,\nonumber\\
c_5 &=& 0.0085,\;\;\;c_6 = -0.0326,\nonumber\\
c_7 &=& 0.0011,\;\;\;c_8 = 0.0004,\nonumber\\
c_9&=& -0.0090,\;\;\;c_{10} = 0.0022.\nonumber\\
\label{2k}
\vspace{2mm}
\end{eqnarray}

One can obtain numerical values of $a_i$ including Wilson coefficients and the color index $N_{c}$\cite{AG}:
\begin{eqnarray}
a_1&=&C_2+C_1/N_{c},\;\; \;a_2=C_1+C_2/N_{c},\nonumber \\
a_3&=&C_3+C_4/N_{c},\;\; \;a_4=C_4+C_3/N_{c},\nonumber \\
a_5&=&C_5+C_6/N_{c},\;\; \;a_6=C_6+C_5/N_{c},\nonumber \\
a_7&=&C_7+C_8/N_{c},\;\; \;a_8=C_8+C_7/N_{c},\nonumber \\
a_9&=&C_9+C_{10}/N_{c},\;\; \;a_{10}= C_{10}+C_{9}/N_{c}.
\label{2k}
\vspace{2mm}
\end{eqnarray}
\section{\label{sec:cpv1}$CP$ asymmetry in $\bar{B}_{s}^{0}\rightarrow \rho^0(\omega)P\rightarrow \pi^+\pi^{-}P$}
\subsection{\label{subsec:form}Formalism}
In the vector meson dominace model (VMD), photons are dressed by coupling to the vector mesons.
Based on the same mechanism, $\rho-\omega$ mixing was proposed and later gradually applied to B
meson physics \cite{Ryoji1996,gar,guo1,guo2,guo11,lei,gang1,gang2,gang3}.
Due to the effective Hamiltonian,
the amplitude $A$ ($\bar{A}$) for the decay process of
$\bar{B}_{s}^{0}\rightarrow\pi^+\pi^{-}P$
(${B}_{s}^{0}\rightarrow\pi^+\pi^{-}\bar{P}$) can be written as \cite{gar}:
\begin{eqnarray}
A=\big<\pi^+\pi^{-}P|H^T|\bar{B}_{s}^{0}\big>+\big<\pi^+\pi^{-}P|H^P|\bar{B}_{s}^{0}\big>,\label{A}
\end{eqnarray}
\begin{eqnarray}
\bar{A}=\big<\pi^+\pi^{-}\bar{P}|H^T|{B}_{s}^{0}\big>+\big<\pi^+\pi^-\bar{P}|H^P|{B}_{s}^{0}\big>,
\end{eqnarray}
with $H^T$ and $H^P$ are the Hamiltonian of the tree and
penguin operators, respectively.

The relative amplitudes and phases of $H^T$ and $H^P$ can be expressed as follows \cite{gar}:
\begin{eqnarray}
A=\big<\pi^+\pi^{-}P|H^T|\bar{B}_{s}^{0}\big>[1+re^{i(\delta+\phi)}],\label{A'}\\
\bar{A}=\big<\pi^+\pi^-\bar{P}|H^T|{B}_{s}^{0}\big>[1+re^{i(\delta-\phi)}],
\label{Abar}
\end{eqnarray}
with $\delta$ and $\phi$ are strong and weak phases, respectively.
$\phi$ is the weak phase in the CKM matrix that causes the CP asymmetry, which
is arg$[V_{tb}V^{*}_{tq}/(V_{ub}V^{*}_{uq})](q=d,s)$. The parameter $r$ represents the
absolute value of the ratio of penguin and tree amplitudes:
\begin{eqnarray}
r\equiv\Bigg|\frac{\big<\pi^+\pi^{-}P|H^P|\bar{B}_{s}^{0}\big>}{\big<\pi^+\pi^{-}P|H^T|\bar{B}_{s}^{0}\big>}\Bigg|
\label{r}.
\end{eqnarray}
The $CP$ violating asymmetry, $A_{CP}$, can be written as
\begin{eqnarray}
A_{CP}\equiv\frac{|A|^2-|\bar{A}|^2}{|A|^2+|\bar{A}|^2}=\frac{-2r
{\rm{sin}}\delta {\rm{sin}}\phi}{1+2r {\rm{cos}}\delta
{\rm{cos}}\phi+r^2}. \label{asy}
\end{eqnarray}
From Equation (\ref{asy}), it can be seen that the $CP$ asymmetry
depends on the weak phase difference and the strong phase
difference. The weak phase is determined for a particular decay process.
Hence, in order to obtain a large $CP$ asymmetry, we
need some mechanism to increase $\delta$. It has been found
that $\rho-\omega$ mixing can lead to
a large strong phase difference
\cite{HB1997,guo1,guo11,lei,guo2,gang1,gang2,gang3}.
Based on $\rho-\omega$ mixing and working to the first order
of isospin violation, we have the following results \cite{gar}:
\begin{eqnarray}
\big<\pi^+\pi^{-}P|H^T|\bar{B}_{s}^{0}\big>=\frac{g_{\rho}}{s_{\rho}s_{\omega}}\widetilde{\Pi}_{\rho\omega}t_{\omega}+\frac{g_{\rho}}{s_{\rho}}t_{\rho},
\label{Htr}\\
\big<\pi^+\pi^{-}P|H^P|\bar{B}_{s}^{0}\big>=\frac{g_{\rho}}{s_{\rho}s_{\omega}}\widetilde{\Pi}_{\rho\omega}p_{\omega}+\frac{g_{\rho}}{s_{\rho}}p_{\rho}.
\label{Hpe}
\end{eqnarray}
where $t_{\rho}(p_{\rho})$ and
$t_{\omega}(p_{\omega})$ are the tree (penguin) amplitudes
for $\bar{B}_{s}^{0}\rightarrow\rho^{0}P$ and
$\bar{B}_{s}^{0}\rightarrow\omega P$, respectively; $g_{\rho}$ is
the coupling constant of $\rho^0\rightarrow\pi^+\pi^-$ decay process;
$\widetilde{\Pi}_{\rho\omega}$ is the effective $\rho-\omega$
mixing amplitude which also effectively absorbed into the direct
coupling $\omega\rightarrow\pi^+\pi^-$. $s_{V}$, $m_{V}$ and $\Gamma_V$($V$=$\rho$ or
$\omega$) represent the inverse propagator, mass and decay rate of the vector meson $V$, respectively.
\begin{eqnarray}
s_V=s-m_V^2+{\rm{i}}m_V\Gamma_V,
\end{eqnarray}
where $\sqrt{s}$ denotes the invariant mass of the $\pi^+\pi^-$
pairs \cite{gar}.

The $\rho-\omega$ mixing paraments were recently determined precisely by Wolfe and Maltnan \cite{CeK,CEK}
\begin{eqnarray}
\mathfrak{Re}{\Pi}_{\rho\omega}(m_{\rho}^2)&=&-4470\pm250_{model}\pm160_{data}
\rm{MeV}^2,\nonumber\\{\mathfrak{Im}}{\Pi}_{\rho\omega}(m_{\rho}^2)&=&-5800\pm2000_{model}\pm1100_ {data}\textrm{MeV}^2
\end{eqnarray}

 One can find that the mixing parameter is the momentum dependence including the non-resonant contribution that absorbs the direct decay $\omega\rightarrow\pi^+\pi^-$.
We introduce the momentum dependence of the mixing parameter $\widetilde{\Pi}_{\rho\omega}(s)$ for $\rho-\omega$ mixing,
which leads to the explicit $s$ dependence. It is reasonable to devote one's energies to search the mixing contribution at the region of $\omega$ mass where the two pions can be produced.
We write $\widetilde{\Pi}_{\rho\omega}(s)={\mathfrak{Re}}\widetilde{\Pi}_{\rho\omega}(m_{\omega}^2)+{\mathfrak{Im}}\widetilde{\Pi}_{\rho\omega}(m_{\omega}^2)$,
and update the values as follows \cite{gang4}:
\begin{eqnarray}
\mathfrak{Re}\widetilde{\Pi}_{\rho\omega}(m_{\omega}^2)&=&-4760\pm440
\rm{MeV}^2,\nonumber\\{\mathfrak{Im}}\widetilde{\Pi}_{\rho\omega}(m_{\omega}^2)&=&-6180\pm3300
\textrm{MeV}^2.
\end{eqnarray}
In fact, the contribution of the $s$ dependence
of $\widetilde{\Pi}_{\rho\omega}$ is negligible. We can make the expansion
$\widetilde{\Pi}_{\rho\omega}(s)=\widetilde{\Pi}_{\rho\omega}(m_{\omega}^2)+(s-m_{\omega})\widetilde{\Pi}_{\rho\omega}^\prime(m_{\omega}^2)$.
From Eqs. (\ref{A})(\ref{A'})(\ref{Htr})(\ref{Hpe}) one has
\begin{eqnarray}
re^{i\delta}e^{i\phi}=\frac{\widetilde{\Pi}_{\rho\omega}p_{\omega}+s_{\omega}p_{\rho}}{\widetilde{\Pi}_{\rho\omega}t_{\omega}+s_{\omega}t_{\rho}},
\label{rdtdirive}
\end{eqnarray}
 Defining
\begin{eqnarray}
\frac{p_{\omega}}{t_{\rho}}\equiv r^\prime
e^{i(\delta_q+\phi)},\quad\frac{t_{\omega}}{t_{\rho}}\equiv
\alpha
e^{i\delta_\alpha},\quad\frac{p_{\rho}}{p_{\omega}}\equiv
\beta e^{i\delta_\beta}, \label{def}
\end{eqnarray}
with $\delta_\alpha$, $\delta_\beta$ and $\delta_q$ are strong
phases. It is available from Eqs.
(\ref{rdtdirive})(\ref{def}):
\begin{eqnarray}
re^{i\delta}=r^\prime
e^{i\delta_q}\frac{\widetilde{\Pi}_{\rho\omega}+\beta
e^{i\delta_\beta}s_{\omega}}{\widetilde{\Pi}_{\rho\omega}\alpha
e^{i\delta_\alpha}+s_{\omega}}. \label{rdt}
\end{eqnarray}
In order to obtain the $CP$ violating asymmetry in Eq.
(\ref{asy}), sin$\phi$ and cos$\phi$ are necessary. The weak phase $\phi$ is
fixed by the CKM matrix elements. In the Wolfenstein
parametrization \cite{wol}, one has
\begin{eqnarray}
{\rm sin}\phi &=&\frac{\eta}{\sqrt{[\rho(1-\rho)-\eta^2]^2+\eta^2}}, \nonumber \\
{\rm cos}\phi &=&\frac{\rho(1-\rho)-\eta^2}{\sqrt{[\rho(1-\rho)-\eta^2]^2+\eta^2}}.
\label{3l1}
\vspace{2mm}
\end{eqnarray}
where the same result has been found for $b\rightarrow d$ transition from $\Lambda_{b}$ decay process \cite{guo1}.

\section{\label{cal}Calculation}
For the simplification, we take the decay process of $\bar B_{s}^0\to\rho^{0}(\omega) K^{0}\rightarrow \pi^{+}\pi^{-} K^{0}$
as example for the study of the $\rho-\omega$ interference.
The other decay channels can be obtained similarly.
According to the Hamiltonian(1), based on CKM matrix elements of $V_{ub}V^{*}_{ud}$,
$V_{tb}V^{*}_{td}$, the decay amplitude of $\bar{B}^{0}_{s}\rightarrow \rho^{0}K^{0}$ in
perturbation QCD approach can be written as
\begin{eqnarray}
\sqrt{2}M(\bar B_{s}^0\to\rho^{0} K^{0})=V_{ub}V^{*}_{ud}t_{\rho}-V_{tb}V^{*}_{td}p_{\rho}
\end{eqnarray}
where $t_{\rho}$ and $p_{\rho}$ refer to the tree and penguin contributions respectively.
We write:
\begin{eqnarray}
t_{\rho}=f_{\rho} F_{B_s\to
K}^{LL} \left[a_{2}\right]+ M_{B_s\to K}^{LL} [C_{2}]
\end{eqnarray}
and
\begin{eqnarray}
p_{\rho}&=&f_{\rho} F_{B_s\to K}^{LL} \left[
 -a_{4}+\frac{3}{2}a_7+\frac{1}{2}a_{10}+\frac{3}{2}a_9\right]
 +  M_{B_s\to K}^{LR} \left[-C_{5}+\frac{1}{2}C_{7}\right]
  \nonumber\\
  &&
   +  M_{B_s\to K}^{LL}
  \left[-C_{3}+\frac{1}{2}C_{9}
  +\frac{3}{2}C_{10}\right]
  -M_{B_s\to K}^{SP} \left[\frac{3}{2}C_{8}\right]
  +  f_{B_s}  F_{ann}^{LL}\left[-a_{4}+ \frac{1}{2}a_{10}\right]
  \nonumber
  \\
  &&
  +  f_{B_s} F_{ann}^{SP}\left[-a_{6}+\frac{1}{2}a_{8}\right]
 +  M_{ann}^{LL}\left[-C_{3}+\frac{1}{2}C_{9}\right]
   +  M_{ann}^{LR}\left[-C_{5}+\frac{1}{2}C_{7}\right]
\end{eqnarray}

The decay amplitude for $\bar{B}^{0}_{s}\rightarrow \omega K^{0}$ can be written as
\begin{eqnarray}
\sqrt{2}M(\bar B_{s}^0\rightarrow \omega\pi^{0})=V_{ub}V^{*}_{ud}t_{\omega}-V_{tb}V^{*}_{td}p_{\omega},
\end{eqnarray}
One can also present the contributions of $t_{\omega}$ and $p_{\omega}$ as well.
\begin{eqnarray}
t_{\omega}=f_\omega F_{B_s\to K}^{LL}
\left[a_{2}\right] + M_{B_s\to K}^{LL} [C_{2}]
\end{eqnarray}
\begin{eqnarray}
p_{\omega}&=&f_\omega F_{B_s\to K}^{LL}  \left[
 2a_{3}+a_{4}+2a_{5}
+\frac{1}{2}a_{7}+\frac{1}{2}a_{9}-\frac{1}{2}a_{10}\right]
\nonumber
   \\
 &&  + M_{B_s\to K}^{LL}\left[C_{3}+2C_{4}-\frac{1}{2}C_9
  +\frac{1}{2}C_{10}\right] +
M_{B_s\to K}^{LR}\left[C_{5}-\frac{1}{2}C_{7}\right]
 \nonumber
  \\
 &&
    - M_{B_s\to K}^{SP}\left[2C_{6}+\frac{1}{2}C_{8}\right]
 + f_{B_s}  F_{ann}^{LL}\left[a_{4} -\frac{1}{2}a_{10}\right]
  +f_{B_s} F_{ann}^{SP}\left[a_{6} -\frac{1}{2}a_{8}\right]
\nonumber
  \\
 &&  + M_{ann}^{LL}\left[C_{3}-\frac{1}{2}C_{9}\right]
 + M_{ann}^{LR}\left[C_{5}-\frac{1}{2}C_{7}\right]
\end{eqnarray}
The function $F$ and $M$ are given in Sec.IX.
The index $LL$, $LR$ and $SP$ arise from the $(V-A)(V-A)$, $(V-A)(V+A)$ and $(S-P)(S+P)$ operators, respectively.
\begin{eqnarray}
\alpha e^{i\delta_\alpha}&=&\frac{t_{\omega}}{t_{\rho}}, \label{eq:afaform} \\
\beta e^{i\delta_\beta}&=&\frac{p_{\rho}}{p_{\omega}}, \label{eq:btaform}
\\
r^\prime e^{i\delta_q}&=&\frac{p_{\omega}}{t_{\rho}}
\times\bigg|\frac{V_{tb}V_{td}^*}{V_{ub}V_{ud}^*}\bigg|,  \label{eq:delform}
\end{eqnarray}
where
\begin{equation}
\left|\frac{V_{tb}V^{*}_{td}}{V_{ub}V^{*}_{ud}}\right|=\frac{\sqrt{[\rho(1-\rho)-\eta^2]^2+\eta^2}}{(1-\lambda^2/2)(\rho^2+\eta^2)}
\label{3p}
\vspace{2mm}
\end{equation}
From above equations, the new strong phases $\delta_\alpha$, $\delta_\beta$ and $\delta_q$ are introduced by the interference
of $\rho-\omega$ mesons. The strong phase $\delta$  are obtained by the equations (\ref{def}) and (\ref{rdt}) in the framework of PQCD.

In a similar way, we can get the $t_{\rho}$, $t_{\omega}$, $p_{\rho}$, and $p_{\omega}$ for the processes of $\bar{B}^{0}_{s}\rightarrow \rho^{0}(\omega) \eta$ and
$\bar{B}^{0}_{s}\rightarrow \rho^{0}(\omega)\eta'$, respectively.
The relevant $CP$ asymmetry can also be produced in similar approach.
In the calculation, $\eta$ and $\eta'$ mesons are introduced.
The $\eta$ and $\eta'$ mixing depend on the quark flavor basis \cite{18}.
The mesons are consisted of $\bar{n}n=(\bar{u}u+\bar{d}d)/\sqrt 2$  and  $\bar{s}s$:
\begin{equation}
\left(\begin{matrix} \big|\eta\big> \\ \big|\eta^{,}\big>
\end{matrix}
\right)=U(\phi)\left(\begin{matrix} \big|\eta_{n}\big> \\ \big|\eta_{s}\big>
\end{matrix}
\right)
=
\left( \begin{array}{ccc}
\cos\phi & -sin\phi \\
\sin\phi  & cos\phi
\end{array} \right)
\left(\begin{matrix} \big|\eta_{n}\big> \\ \big|\eta_{s}\big>
\end{matrix}
\right)
\end{equation}

where the mixing angle $\phi=39.3^{\circ}\pm1.0^{\circ}$.  Explicitly, only two decay constants are needed is the advantage here:
\begin{eqnarray}
\big<0|\bar{n}\gamma^{\mu}\gamma_{5}n|\eta_{n}P^{\mu}\big>=\frac{i}{\sqrt2}{f_{n}P^{\mu}},
\label{gatt}\\
\big<0|\bar{s}\gamma^{\mu}\gamma_{5}s|\eta_{s}P^{\mu}\big>=if_{s}P^{\mu}.
\label{gat2}
\end{eqnarray}
We use \cite{19}
\begin{eqnarray}
f_{n}=139.1\pm2.6 MeV, && f_{s}=174.2\pm7.8 MeV.
\label{fft}
\end{eqnarray}

For the pure annihilation type decay process, one can also divides the amplitudes into $t_{\rho}$, $t_{\omega}$, $p_{\rho}$, and $p_{\omega}$ depending on
$V_{ub}V^{*}_{us}$ and $V_{tb}V^{*}_{ts}$. The amplitudes can be given as following for the channel $\bar{B}^{0}_{s}\rightarrow \pi^{0}\rho^{0}(\omega)$:
$M(\bar B_{s}^0\to\rho^{0}\pi^{0})=V_{ub}V^{*}_{us}t_{\rho}-V_{tb}V^{*}_{ts}p_{\rho}$ and $M(\bar B_{s}^0\to \omega\pi^{0})=V_{ub}V^{*}_{us}t_{\omega}-V_{tb}V^{*}_{ts}p_{\omega}$.

\section{\label{input}Input parameters}

The CKM matrix, which elements are determined from experiments, can be expressed in terms of the Wolfenstein parameters $A$, $\rho$, $\lambda$ and $\eta$ \cite{wol}:
\begin{equation}
\left(
\begin{array}{ccc}
  1-\tfrac{1}{2}\lambda^2   & \lambda                  &A\lambda^3(\rho-\mathrm{i}\eta) \\
  -\lambda                 & 1-\tfrac{1}{2}\lambda^2   &A\lambda^2 \\
  A\lambda^3(1-\rho-\mathrm{i}\eta) & -A\lambda^2              &1\\
\end{array}
\right),\label{ckm}
\end{equation}
where $\mathcal{O} (\lambda^{4})$ corrections are neglected. The latest values for the parameters in the CKM matrix are \cite{PDG2016}:
\begin{eqnarray}
&& \lambda=0.22506\pm0.00050,\quad A=0.811\pm 0.026,\nonumber \\
&& \bar{\rho}=0.124_{-0.018}^{+0.019},\quad
\bar{\eta}=0.356\pm{0.011}.\label{eq: rhobarvalue}
\end{eqnarray}
where
\begin{eqnarray}
 \bar{\rho}=\rho(1-\frac{\lambda^2}{2}),\quad
\bar{\eta}=\eta(1-\frac{\lambda^2}{2}).\label{eq: rho rhobar
relation}
\end{eqnarray}
From Eqs. (\ref{eq: rhobarvalue}) ( \ref{eq: rho rhobar relation})
we have
\begin{eqnarray}
0.109<\rho<0.147,\quad  0.354<\eta<0.377.\label{eq: rho value}
\end{eqnarray}
The other parameters are given as following \cite{wol,ADR2016,XZZ2016}:
\begin{eqnarray}
f_{\pi}&=&0.131 \text{GeV}, \hspace{2.95cm}  f_{K}=0.160 \text{GeV},        \nonumber \\
m_{B^0_s}&=&5.36677 \text{GeV},\hspace{2.95cm} \tau_{B^0_s}=1.512\times10^{-12} s   \nonumber \\
m_{\rho^0(770)}&=&0.77526 \text{GeV}, \hspace{2.95cm} \Gamma_{\rho^0(770)}=0.1491 \text{GeV},  \nonumber \\
m_{\omega(782)}&=&0.78265 \text{GeV}, \hspace{2.95cm} \Gamma_{\omega(782)}=8.49\times10^{-3} \text{GeV},  \nonumber \\
m_\pi&=&0.13957 \text{GeV},\hspace{2.95cm} m_W=80.385 \text{GeV},  \nonumber \\
f_\rho&=&209\pm2 \text{MeV},\hspace{2.95cm} f_\rho^T=165\pm9 \text{MeV},  \nonumber \\
f_\omega&=&195.1\pm3 \text{MeV},\hspace{2.95cm} f_\omega^T=145\pm10 \text{MeV}.
\end{eqnarray}

\section{\label{sec:numerical}Numerical results}

In the numerical results, we find the $CP$ asymmetry can be enhanced when the masses of the $\pi^+\pi^-$ pairs are in the area around the $\rho-\omega$
resonance, and the maximum $CP$ asymmetry for our considering the decay channels can reach 59{\%}. We also discuss the numerical results from the case of tree and penguin
dominated type decay and the case of pure annihilation type decay in the framework of Perturbative QCD. The CP violation is associated with the CKM matrix elements and $\sqrt{s}$.
In our numerical calculations, we find that the CP asymmetry depend weakly on the variation of the CKM matrix elements. Hence, we
let $(\rho,\eta)$ vary between the central values $(\rho_{central},\eta_{central})$.

\subsection{\label{num} The case of tree and penguin dominated type decay}

\begin{figure}
\resizebox{0.5\textwidth}{!}{%
  \includegraphics{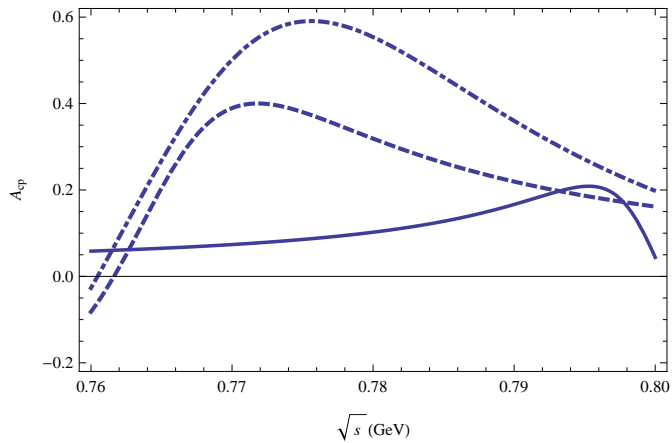}
  }
\caption{Plot of $A_{CP}$ as a function of $\sqrt{s}$
corresponding to central parameter values of CKM matrix elements. The dashed line, dash-dotted, solid line refer to the decay channels of $\bar{B}^{0}_{s}\rightarrow \rho^{0}(\omega)K^{0}\rightarrow \pi^{+}\pi^{-}K^{0}$ ,
 $\bar{B}^{0}_{s}\rightarrow \rho^{0}(\omega)\eta'\rightarrow \pi^{+}\pi^{-}\eta' $ and $\bar{B}^{0}_{s}\rightarrow \rho^{0}(\omega)\eta\rightarrow \pi^{+}\pi^{-}\eta $, respectively.
}
\label{fig:1}       
\end{figure}

\begin{figure}
\resizebox{0.5\textwidth}{!}{%
  \includegraphics{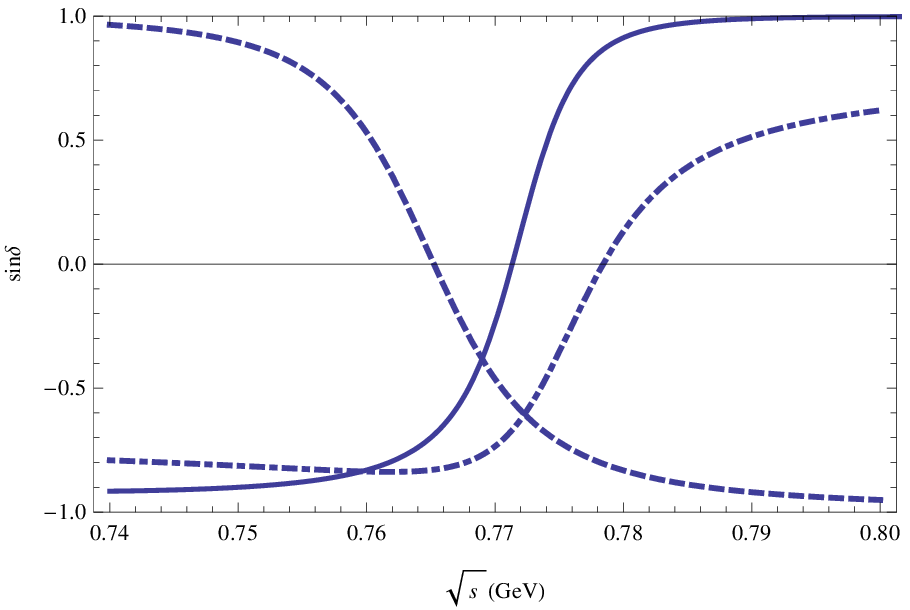}
  }
\caption{Plot of $\sin\delta$ as a function of $\sqrt{s}$
corresponding to central parameter values of CKM matrix elements. The dashed line, dash-dotted, solid line refer to the decay channels of $\bar{B}^{0}_{s}\rightarrow \rho^{0}(\omega)K^{0}\rightarrow \pi^{+}\pi^{-}K^{0}$ ,
 $\bar{B}^{0}_{s}\rightarrow \rho^{0}(\omega)\eta'\rightarrow \pi^{+}\pi^{-}\eta' $ and $\bar{B}^{0}_{s}\rightarrow \rho^{0}(\omega)\eta\rightarrow \pi^{+}\pi^{-}\eta $, respectively.)
}
\label{fig:2}       
\end{figure}

\begin{figure}
\resizebox{0.5\textwidth}{!}{%
  \includegraphics{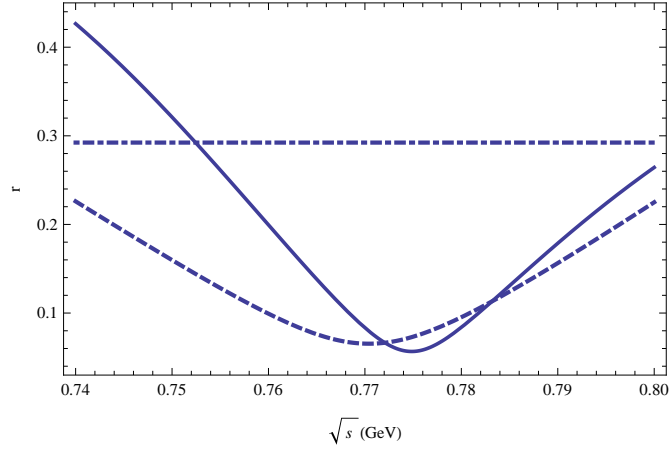}
  }
\caption{Plot of r as a function of $\sqrt{s}$
corresponding to central parameter values of CKM matrix elements. The dashed line, dash-dotted, solid line refer to the decay channels of $\bar{B}^{0}_{s}\rightarrow \rho^{0}(\omega)K^{0}\rightarrow \pi^{+}\pi^{-}K^{0}$ ,
 $\bar{B}^{0}_{s}\rightarrow \rho^{0}(\omega)\eta'\rightarrow \pi^{+}\pi^{-}\eta' $ and $\bar{B}^{0}_{s}\rightarrow \rho^{0}(\omega)\eta\rightarrow \pi^{+}\pi^{-}\eta $, respectively.)
}
\label{fig:3}       
\end{figure}

We refer to the decay processes of $\bar{B}^{0}_{s}\rightarrow \rho^{0}(\omega)K^{0}\rightarrow \pi^{+}\pi^{-}K^{0}$, $\bar{B}^{0}_{s}\rightarrow \rho^{0}(\omega) \eta\rightarrow \pi^{+}\pi^{-} \eta$ and $\bar{B}^{0}_{s}\rightarrow \rho^{0}(\omega)\eta' \rightarrow \pi^{+}\pi^{-}\eta'$ as the case of tree and penguin dominated type decay. In Fig.1, we show the plot of CP asymmetry as a function of $\sqrt{s}$.
One can find the $CP$ asymmetry varies sharply when the masses of the $\pi^{+}\pi^{-}$ pairs are in the area around the $\rho-\omega$
resonance range. For the decay process of $\bar{B}^{0}_{s}\rightarrow \rho^{0}(\omega)K^{0}\rightarrow \pi^{+}\pi^{-}K^{0} $ , the maximum CP asymmetry can reach $40\%$.
For the decay channels of $\bar{B}^{0}_{s}\rightarrow \rho^{0}(\omega)\eta'\rightarrow \pi^{+}\pi^{-}\eta' $ and $\bar{B}^{0}_{s}\rightarrow \rho^{0}(\omega)\eta\rightarrow \pi^{+}\pi^{-}\eta $,
we obtain the maximum CP asymmetry is $59\%$ and $21\%$, respectively. From Equation (\ref{asy}), one can find the $CP$ asymmetry
is affected by the weak phase difference, the strong phase
difference and $r$. The weak phase depends on the CKM matrix elements.
Hence, the change of $CP$ asymmetry is derived from the variation of strong phase $\delta$ and $r$ except the CKM matrix.
We take the central values from the parameters of $(\rho_{central},\eta_{central})$.
Taking into account of $\rho-\omega$ mixing, we can see that $\sin\delta$ oscillate considerably at the area of $\rho-\omega$ resonance from Fig.2 for the considering decay processes.
The plot of r as a function of $\sqrt{s}$ is presented in Fig.3.  One can see that the $r$ change sharply for the process of $\bar{B}^{0}_{s}\rightarrow \rho^{0}(\omega)\eta'\rightarrow \pi^{+}\pi^{-}\eta'$ and $\bar{B}^{0}_{s}\rightarrow \rho^{0}(\omega)\eta\rightarrow \pi^{+}\pi^{-}\eta $.

\subsection{\label{num}The case of pure annihilation decay type}

\begin{figure}
\resizebox{0.5\textwidth}{!}{%
  \includegraphics{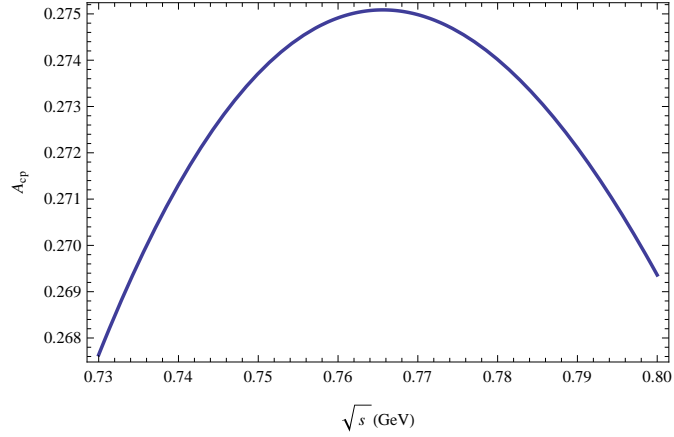}
  }
\caption{Plot of $A_{CP}$ as a function of $\sqrt{s}$
corresponding to central parameter values of CKM matrix elements
for  $\bar{B}^{0}_{s}\rightarrow \rho^{0}(\omega)\pi^{0}\rightarrow \pi^{+}\pi^{-}\pi^{0}$.
}
\label{fig:4}       
\end{figure}
\begin{figure}
\resizebox{0.5\textwidth}{!}{%
  \includegraphics{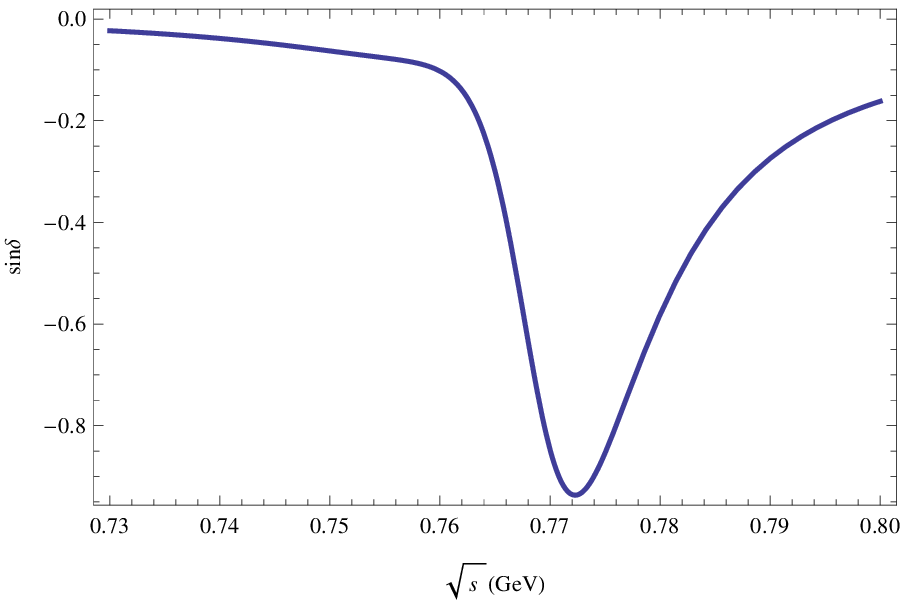}
  }
\caption{Plot of $\sin\delta$ as a function of $\sqrt{s}$
corresponding to central parameter values of CKM matrix elements
for $\bar{B}^{0} _{s}\rightarrow \rho^{0}(\omega)\pi^{0}\rightarrow \pi^{+}\pi^{-}\pi^{0}$.}
\label{fig:5}       
\end{figure}

\begin{figure}
\resizebox{0.5\textwidth}{!}{%
  \includegraphics{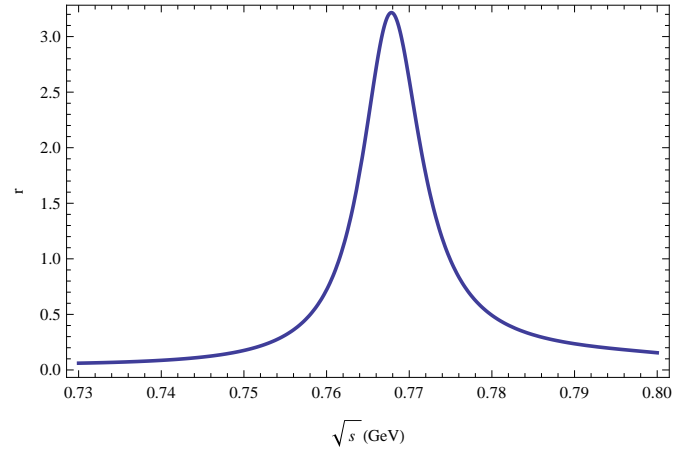}
  }
\caption{Plot of r as a function of $\sqrt{s}$
corresponding to central parameter values of CKM matrix elements
for  $\bar{B}^{0}_{s}\rightarrow \rho^{0}(\omega)\pi^{0}\rightarrow \pi^{+}\pi^{-}\pi^{0}$.
}
\label{fig:6}       
\end{figure}

In Fig.4, we present the plot of CP asymmetry parameter as a function $\sqrt{s}$
corresponding to central parameter values of CKM matrix elements
for the pure annihilation decay type of $\bar{B}^{0}_{s}\rightarrow \rho^{0}(\omega)\pi^{0}\rightarrow \pi^{+}\pi^{-}\pi^{0}$.
One can find the maximum CP asymmetry reach $28\%$ when the masses of the $\pi^{+}\pi^{-}$ pairs are in the area around the $\rho-\omega$
resonance range. The plots of $\sin\delta$ and $r$ as a function of $\sqrt{s}$ are given in Fig.5 and Fig.6, respectively.
We can see that $\sin\delta$ and $r$ oscillate sharply taking into account $\rho-\omega$ resonance.
Generally, the $CP$ asymmetry is tiny in the case of pure annihilation decay.
However, the maximum $CP$ asymmetry can reach $28\%$ at the area of $\rho-\omega$ resonance, which give us a
chance to search $CP$ asymmetry from the pure annihilation decay type.

\section{\label{sec:conclusion}Summary and conclusion}

In this paper, we study the $CP$ asymmetry for the decay process of $\bar{B}_{s}\rightarrow P\pi^+\pi^-$ in Perturbative QCD.
It has been found the $CP$ asymmetry can be enhanced greatly at the area of $\rho-\omega$ resonance.
The maximum $CP$ asymmetry can reach $40\%$ for the process of $\bar{B}^{0}_{s}\rightarrow \rho^{0}(\omega)K^{0}\rightarrow \pi^{+}\pi^{-}K^{0}$.
However, the paper has also discussed the $CP$ asymmetry of the decay process of $\bar{B}_{s}\rightarrow\rho^{0}(\omega)K ^{0}
\rightarrow\pi^{+}\pi^{-} K ^{0}$ from $b \rightarrow d$ transition
in QCD factorization.
The maximum $CP$ asymmetry reach $46\%$ when the invariant mass of the
$\pi^{+}\pi^{-}$ pair is in the vicinity of the $\omega$ resonance from QCD factorization \cite{gang2}.
The difference of $CP$ asymmetry mainly comes from the strong phase difference between
QCD factoriztion and  Perturbative QCD. The hadronic matrix elements can be calculated
from first principles in the decays of B-meson.
Due to the power expansion of $1/m_b$ ($m_b$ is b quark mass),
all of the theories of factorization are shown to deal with
the hadronic matrix elements in the leading power of $1/m_b$.
But these methods are different significantly due to the collinear degree
or transverse momenta. The power counting is different from the hard kernels
between QCDF and PQCD. It is important to extract the strong phase difference
for $CP$ violation. The more different feature of QCDF and PQCD is
the strong interaction scale at which of PQCD is low, typically of order
$1\sim 2 $ GeV, the case of QCDF is order $O(m_b)$ for the Wilson coefficients.

Meanwhile, we find that the $CP$ asymmetry associated with the case of pure annihilation type decay process of
$\bar{B}^{0}_{s}\rightarrow \rho^{0}(\omega)\pi^{0}\rightarrow \pi^{+}\pi^{-}\pi^{0}$ can be enhanced and the maximum value reach $28\%$.
Hence, one can search for the large $CP$ asymmetry at the area of $\rho-\omega$ resonance from pure annihilation type decay process of
$\bar{B}^{0}_{s}\rightarrow \rho^{0}(\omega)\pi^{0}\rightarrow \pi^{+}\pi^{-}\pi^{0}$.

In this work, we have take the Perturbative QCD approximation which add the QCD correction to the naive factorization which is based
on the power expansion of $1/m_{b}$. The final state interaction is also neglected in this
approximation which may give some uncertainties.  There are some uncertainties from the input parameters, the hard scattering scale and
CKM matrix elements. The theoretical results can be improved by high order correction from  $\alpha_{s}$ and $1/m_{b}$.

\section{Acknowledgments}
This work was supported by National Natural Science
Foundation of China (Project Numbers 11605041), Plan For Scientific Innovation Talent of Henan University of Technology
 (Project Number 2012CXRC17), the Key Project (Project Number 14A140001)
 for Science and Technology of the Education Department Henan Province,
the Fundamental Research Funds (Project Number 2014YWQN06) for the Henan Provincial Colleges and Universities,
and the Research Foundation of the young core teacher from Henan province.

\section{APPENDIX: Related functions defined in the text}
The functions associated with the tree and penguin contributions are presented for the factorization and non-factorization
amplitudes in PQCD approach \cite{YA,LKM,AMLi2007}. The functions of the case of tree and penguin dominated type decay are written as
\begin{itemize}
\item $(V-A)(V-A)$ operators:
  \begin{eqnarray}
  f_{M_2} F^{LL}_{B_s\to M_3} (a_i)&=&8\pi
  C_FM_{B_s}^4f_{M_2}\int^1_0dx_1dx_3\int^\infty_0b_1db_1b_3db_3
\phi_{B_s}(x_1,b_1)
  \Big\{a_i(t_a) E_e(t_a)
  \nonumber\\
  &&\times \Big[(1+x_3)\phi_3^A(x_3)+r_3(1-2x_3)(\phi_3^P(x_3)+\phi_3^T(x_3))
  \Big]h_e(x_1,x_3,b_1,b_3)
 \nonumber\\ && \;\;+2r_3\phi_3^P(x_3)a_i(t_a^\prime) E_e(t_a^\prime)h_e(x_3,x_1,b_3,b_1)
 \Big\},\label{ppefll}
  \end{eqnarray}

\item $(V-A)(V+A)$ operators:
\begin{eqnarray}
  F^{LR}_{B_s\to M_3}(a_i)&=&-F^{LL}_{B_s\to M_3}(a_i),\label{ppeflr}
\end{eqnarray}

\item $(S-P)(S+P)$ operators:
  \begin{eqnarray}
 f_{M_2} F^{SP}_{B_s\to M_3}(a_i)&=& 16\pi r_2
  C_FM_{B_s}^4f_{M_2}\int^1_0dx_1dx_3\int^\infty_0b_1db_1b_3db_3
\phi_{B_s}(x_1,b_1)
  \Big\{a_i(t_a)E_e(t_a)\nonumber\\
  &&\times\Big[\phi_3^A(x_3)+r_3(2+x_3)\phi_3^P(x_3)-r_3x_3\phi_3^T(x_3)\Big]
  h_e(x_1,x_3,b_1,b_3)\nonumber\\
  &&\;\;\;+2
  r_3\phi_3^P(x_3)a_i(t^\prime_a)E_e(t_a^\prime)h_e(x_3,x_1,b_3,b_1)\Big\},\label{ppefsp}
  \end{eqnarray}
\end{itemize}

\begin{itemize}
\item $(V-A)(V-A)$ operators:
\begin{eqnarray} M_{B_s\to M_3}^{LL}(a_i)&=&32\pi
C_FM_{B_s}^4/\sqrt{6}\int^1_0dx_1dx_2dx_3\int^\infty_0b_1db_1b_2db_2
\phi_{B_s}(x_1,b_1)\phi_2^A(x_2)
\nonumber\\
&&\times
\Big\{\Big[(1-x_2)\phi_3^A(x_3)-r_3x_3(\phi_3^P(x_3)-\phi_3^T(x_3))\Big]
a_i(t_b)E_e^\prime(t_b)\nonumber\\
&&~\times h_n(x_1,1-x_2,x_3,b_1,b_2)+h_n(x_1,x_2,x_3,b_1,b_2)\nonumber\\
 &&\;\;\times\Big[-(x_2+x_3)\phi_3^A(x_3)+r_3x_3(\phi_3^P(x_3)+\phi_3^T(x_3))\Big]
 a_i(t_b^\prime) E_e^\prime(t_b^\prime)\Big\},\label{ppenll}
 \end{eqnarray}

\item $(V-A)(V+A)$ operators:
\begin{eqnarray} M_{B_s\to M_3}^{LR}( a_i)&=&32\pi C_FM_{B_s}^4r_2/\sqrt{6}
      \int^1_0dx_1dx_2dx_3\int^\infty_0b_1db_1b_2db_2\phi_{B_s}(x_1,b_1)\nonumber\\
     &&\times \Big\{h_n(x_1,1-x_2,x_3,b_1,b_2)\Big[(1-x_2)\phi_3^A(x_3)
     \left(\phi_2^P(x_2)+\phi_2^T(x_2)\right)\nonumber\\
     &&\;\;+r_3x_3\left(\phi_2^P(x_2)-\phi_2^T(x_2)\right)
     \left(\phi_3^P(x_3)+\phi_3^T(x_3)\right)\nonumber\\
     &&\;\;+(1-x_2)r_3\left(\phi_2^P(x_2)+\phi_2^T(x_2)\right)\left(\phi_3^P(x_3)
      -\phi_3^T(x_3)\right)\Big]a_i(t_b)
         E_e^\prime(t_b) \nonumber\\
       &&\;\;-h_n(x_1,x_2,x_3,b_1,b_2)\Big[x_2\phi_3^A(x_3)(\phi_2^P(x_2)-\phi_2^T(x_2))\nonumber\\
       &&\;\;+r_3x_2(\phi_2^P(x_2)-\phi_2^T(x_2))(\phi_3^P(x_3)-\phi_3^T(x_3))\nonumber\\
      &&\;\;+r_3x_3(\phi_2^P(x_2)+\phi_2^T(x_2))(\phi_3^P(x_3)+\phi_3^T(x_3))\Big]a_i(t^\prime_b)
       E_e^\prime(t_b^\prime)\Big\},\label{ppenlr}
 \end{eqnarray}

\item $(S-P)(S+P)$ operators:
\begin{eqnarray} M^{SP}_{B_s\to M_3}( a_i) &=&32\pi C_F
M_{B_s}^4/\sqrt{6}\int^1_0dx_1dx_2dx_3\int^\infty_0b_1db_1b_2db_2
\phi_{B_s}(x_1,b_1)\phi_2^A(x_2)
\nonumber\\
&&\times\Big\{
\Big[(x_2-x_3-1)\phi_3^A(x_3)+r_3x_3(\phi_3^P(x_3)+\phi_3^T(x_3))\Big]\nonumber\\
&&\times
a_i(t_b)E_e^\prime(t_b)h_n(x_1,1-x_2,x_3,b_1,b_2)+a_i(t_b^\prime)
E^\prime_e(t_b^\prime)\nonumber\\
&&\times
 \Big[x_2\phi_3^A(x_3)+r_3x_3(\phi_3^T(x_3)-\phi_3^P(x_3))\Big]h_n(x_1,x_2,x_3,b_1,b_2)\Big\}.
 \label{ppensp}
\end{eqnarray}
\end{itemize}

The functions are associated with the annihilation type process as following:
\begin{itemize}
\item $(V-A)(V-A)$ operators:
\begin{eqnarray}
f_{B_s} F_{ann}^{LL}( a_i)&=&8\pi
C_FM_{B_s}^4f_{B_s}\int^1_0dx_2dx_3\int^\infty_0b_2db_2b_3db_3\Big\{a_i(t_c)
E_a(t_c)
\nonumber\\
&&
\times\Big[(x_3-1)\phi_2^A(x_2)\phi_3^A(x_3)-4r_2r_3\phi_2^P(x_2)\phi_3^P(x_3)\nonumber
\\
&&+2r_2r_3x_3\phi_2^P(x_2)(\phi_3^P(x_3)-\phi_3^T(x_3))\Big]h_a(x_2,1-x_3,b_2,b_3)\nonumber
\\
&&+\Big[x_2\phi_2^A(x_2)
\phi_3^A(x_3)+2r_2r_3(\phi_2^P(x_2)-\phi_2^T(x_2))\phi_3^P(x_3)\nonumber\\
&&+2r_2r_3x_2(\phi_2^P(x_2)+\phi_2^T(x_2))\phi_3^P(x_3)\Big]
a_i(t_c^\prime)
E_a(t_c^\prime)h_a(1-x_3,x_2,b_3,b_2)\Big\}.\label{ppafll}
 \end{eqnarray}

 \item
$(V-A)(V+A)$ operators:
\begin{eqnarray}
F_{ann}^{LR}( a_i)=F_{ann}^{LL}(a_i),\label{ppaflr}
\end{eqnarray}

 \item $(S-P)(S+P)$ operators:
 \begin{eqnarray}
 f_{B_s} F_{ann}^{SP}(a_i)&=&16\pi
  C_FM_{B_s}^4f_{B_s}\int^1_0dx_2dx_3\int^\infty_0b_2db_2b_3db_3
  \Big\{\Big[2r_2\phi_2^P(x_2)\phi_3^A(x_3)\nonumber\\
  &&\;\;+(1-x_3)r_3\phi_2^A(x_2)(\phi_3^P(x_3)
  +\phi_3^T(x_3))\Big]
 a_i(t_c) E_a(t_c)h_a(x_2,1-x_3,b_2,b_3)\nonumber\\
  &&\;\;+\Big[2r_3\phi_2^A(x_2)\phi_3^P(x_3)+r_2x_2(\phi_2^P(x_2)-\phi_2^T(x_2))\phi_3^A(x_3)
  \Big]\nonumber\\
  &&\;\;\times
  a_i(t_c^\prime)E_a(t_c^\prime)h_a(1-x_3,x_2,b_3,b_2)\Big\}.\label{ppafsp}
\end{eqnarray}
\end{itemize}

\begin{itemize}
\item $(V-A)(V-A)$ operators:
\begin{eqnarray}
 M_{ann}^{LL}( a_i)&=&32\pi C_FM_{B_s}^4/\sqrt
 {6}\int^1_0dx_1dx_2dx_3\int^\infty_0b_1db_2b_2db_2\phi_{B_s}(x_1,b_1)\nonumber\\
 &&\times \Big\{h_{na}(x_1,x_2,x_3,b_1,b_2)\Big[-x_2\phi_2^A(x_2)\phi_3^A(x_3)-4r_2r_3
 \phi_2^P(x_2)\phi_3^P(x_3)\nonumber\\
 &&\;\;\;+r_2r_3(1-x_2)(\phi_2^P(x_2)+\phi_2^T(x_2))(\phi_3^P(x_3)-\phi_3^T(x_3))
 \nonumber\\
 &&\;\;+r_2r_3x_3(\phi_2^P(x_2)-\phi_2^T(x_2))(\phi_3^P(x_3)+\phi_3^T(x_3))\Big]a_i(t_d)
 E_a^\prime(t_d)\nonumber\\
 &&\;\;+h_{na}^\prime(x_1,x_2,x_3,b_1,b_2)\Big[(1-x_3)\phi_2^A(x_2)\phi_3^A(x_3)
 \nonumber\\
 &&\;\;+(1-x_3)r_2r_3(\phi_2^P(x_2)+\phi_2^T(x_2))(\phi_3^P(x_3)-\phi_3^T(x_3))
 \nonumber\\
 &&\;\;+x_2r_2r_3(\phi_2^P(x_2)-\phi_2^T(x_2))(\phi_3^P(x_3)+\phi_3^T(x_3))\Big]
 a_i(t_d^\prime)
 E_a^\prime(t_d^\prime)\Big\},\label{ppanll}
 \end{eqnarray}

 \item $(V-A)(V+A)$ operators:
 \begin{eqnarray}
 M_{ann}^{LR}(M_2,M_3, a_i)&=&32\pi C_FM_{B_s}^4/\sqrt
 {6}\int^1_0dx_1dx_2dx_3\int^\infty b_1db_1b_2db_2\phi_{B_s}(x_1,b_1)\nonumber\\
 &&\;\;\times\Big\{h_{na}(x_1,x_2,x_3,b_1,b_2)\Big[r_2(2-x_2)(\phi_2^P(x_2)+\phi_2^T(x_2))
 \phi_3^A(x_3)\nonumber\\
 &&\;\;-r_3(1+x_3)\phi_2^A(x_2)(\phi_3^P(x_3)-\phi_3^T(x_3))\Big]a_i(t_d)E_a^\prime(t_d)
 \nonumber\\
 &&\;\;+h_{na}^\prime
 (x_1,x_2,x_3,b_1,b_2)\Big[r_2x_2\left(\phi_2^P(x_2)+\phi_2^T(x_2)\right)\phi_3^A(x_3)
 \nonumber\\
 &&\;\;+r_3(x_3-1)\phi_2^A(x_2)
 (\phi_3^P(x_3)-\phi_3^T(x_3))\Big]  a_i(t_d^\prime)E_a^\prime(t_d^\prime)
 \Big\},\label{ppanlr}
 \end{eqnarray}

 \item $(S-P)(S+P)$ operators:
 \begin{eqnarray}
 M_{ann}^{SP}( a_i)&=&32\pi C_F M_{B_s}^4/\sqrt {6}\int^1_0dx_1dx_2dx_3\int^\infty_0b_1db_1b_2db_2
 \phi_{B_s}(x_1,b_1)\nonumber\\
 &&\times \Big\{a_i(t_d)E_a^\prime(t_d)h_{na}(x_1,x_2,x_3,b_1,b_2)\Big[(x_3-1)
 \phi_2^A(x_2)\phi_3^A(x_3)\nonumber\\
 &&\;\; -4r_2r_3\phi_2^P(x_2)\phi_3^P(x_3)+r_2r_3x_3(\phi_2^P(x_2)+\phi_2^T(x_2))
 (\phi^P_3(x_3)-\phi_3^T(x_3))\nonumber\\
 &&\;\;+r_2r_3(1-x_2)(\phi_2^P(x_2)-\phi_2^T(x_2))(\phi^P_3(x_3)+\phi_3^T(x_3))\Big]
 \nonumber\\
 &&\;\;+a_i(t_d^\prime)
 E_a^\prime(t_d^\prime)h_{na}^\prime(x_1,x_2,x_3,b_1,b_2)
  \Big[x_2\phi_2^A(x_2)\phi_3^A(x_3)\nonumber
 \\
 &&\;\;+x_2r_2r_3(\phi_2^P(x_2)+\phi_2^T(x_2))
 (\phi_3^P(x_3)-\phi_3^T(x_3)))\nonumber\\
 &&\;\;+r_2r_3(1-x_3)(\phi_2^P(x_2)-\phi_2^T(x_2))(\phi_3^P(x_3)+\phi_3^T(x_3))\Big]\Big\}.
 \label{ppansp}
 \end{eqnarray}
\end{itemize}

The hard scales are chosen as \begin{eqnarray}
t_a&=&\mbox{max}\{{\sqrt
{x_3}M_{B_s},1/b_1,1/b_3}\},\\
t_a^\prime&=&\mbox{max}\{{\sqrt
{x_1}M_{B_s},1/b_1,1/b_3}\},\\
t_b&=&\mbox{max}\{\sqrt
{x_1x_3}M_{B_s},\sqrt{|1-x_1-x_2|x_3}M_{B_s},1/b_1,1/b_2\},\\
t_b^\prime&=&\mbox{max}\{\sqrt{x_1x_3}M_{B_s},\sqrt
{|x_1-x_2|x_3}M_{B_s},1/b_1,1/b_2\},\\
t_c&=&\mbox{max}\{\sqrt{1-x_3}M_{B_s},1/b_2,1/b_3\},\\
t_c^\prime
&=&\mbox{max}\{\sqrt {x_2}M_{B_s},1/b_2,1/b_3\},\\
t_d&=&\mbox{max}\{\sqrt {x_2(1-x_3)}M_{B_s},
\sqrt{1-(1-x_1-x_2)x_3}M_{B_s},1/b_1,1/b_2\},\\
t_d^\prime&=&\mbox{max}\{\sqrt{x_2(1-x_3)}M_{B_s},\sqrt{|x_1-x_2|(1-x_3)}M_{B_s},1/b_1,1/b_2\}.
\end{eqnarray}

The functions $h$  in the decay amplitudes consist of two parts: one
is the jet function $S_t(x_i)$ derived by the threshold
re-summation\cite{L3}, the other is the propagator of virtual quark
and gluon. They are defined by
\begin{eqnarray}
h_e(x_1,x_3,b_1,b_3)&=&\left[\theta(b_1-b_3)I_0(\sqrt
x_3M_{B_s}b_3)K_0(\sqrt
x_3 M_{B_s}b_1)\right.\\
&& \left.+\theta(b_3-b_1)I_0(\sqrt x_3M_{B_s}b_1)K_0(\sqrt
x_3M_{B_s}b_3)\right]K_0(\sqrt {x_1x_3}M_{B_s}b_1)S_t(x_3),\nonumber\\
h_n(x_1,x_2,x_3,b_1,b_2)&=&\left[\theta(b_2-b_1)K_0(\sqrt
{x_1x_3}M_{B_s}b_2)I_0(\sqrt
{x_1x_3}M_{B_s}b_1)\right. \nonumber\\
&&\;\;\;\left. +\theta(b_1-b_2)K_0(\sqrt
{x_1x_3}M_{B_s}b_1)I_0(\sqrt{x_1x_3}M_{B_s}b_2)\right]\nonumber\\
&&\times
\left\{\begin{array}{ll}\frac{i\pi}{2}H_0^{(1)}(\sqrt{(x_2-x_1)x_3}
M_{B_s}b_2),& x_1-x_2<0\\
K_0(\sqrt{(x_1-x_2)x_3}M_{B_s}b_2),& x_1-x_2>0
\end{array}
\right. ,
\end{eqnarray}
\begin{eqnarray}
h_a(x_2,x_3,b_2,b_3)&=&(\frac{i\pi}{2})^2
S_t(x_3)\Big[\theta(b_2-b_3)H_0^{(1)}(\sqrt{x_3}M_{B_s}b_2)J_0(\sqrt
{x_3}M_{B_s}b_3)\nonumber\\
&&\;\;+\theta(b_3-b_2)H_0^{(1)}(\sqrt {x_3}M_{B_s}b_3)J_0(\sqrt
{x_3}M_{B_s}b_2)\Big]H_0^{(1)}(\sqrt{x_2x_3}M_{B_s}b_2),
\\
h_{na}(x_1,x_2,x_3,b_1,b_2)&=&\frac{i\pi}{2}\left[\theta(b_1-b_2)H^{(1)}_0(\sqrt
{x_2(1-x_3)}M_{B_s}b_1)J_0(\sqrt {x_2(1-x_3)}M_{B_s}b_2)\right. \nonumber\\
&&\;\;\left.
+\theta(b_2-b_1)H^{(1)}_0(\sqrt{x_2(1-x_3)}M_{B_s}b_2)J_0(\sqrt
{x_2(1-x_3)}M_{B_s}b_1)\right]\nonumber\\
&&\;\;\;\times K_0(\sqrt{1-(1-x_1-x_2)x_3}M_{B_s}b_1),
\\
h_{na}^\prime(x_1,x_2,x_3,b_1,b_2)&=&\frac{i\pi}{2}\left[\theta(b_1-b_2)H^{(1)}_0(\sqrt
{x_2(1-x_3)}M_{B_s}b_1)J_0(\sqrt{x_2(1-x_3)}M_{B_s}b_2)\right. \nonumber\\
&&\;\;\;\left. +\theta(b_2-b_1)H^{(1)}_0(\sqrt
{x_2(1-x_3)}M_{B_s}b_2)J_0(\sqrt{x_2(1-x_3)}M_{B_s}b_1)\right]\nonumber\\
&&\;\;\;\times
\left\{\begin{array}{ll}\frac{i\pi}{2}H^{(1)}_0(\sqrt{(x_2-x_1)(1-x_3)}M_{B_s}b_1),&
x_1-x_2<0\\
K_0(\sqrt {(x_1-x_2)(1-x_3)}M_{B_s}b_1),&
x_1-x_2>0\end{array}\right. ,
\end{eqnarray}
where $H_0^{(1)}(z) = \mathrm{J}_0(z) + i\, \mathrm{Y}_0(z)$.

The $S_t$ re-sums the threshold logarithms $\ln^2x$ appearing in the
hard kernels to all orders and it has been parameterized as
  \begin{eqnarray}
S_t(x)=\frac{2^{1+2c}\Gamma(3/2+c)}{\sqrt \pi
\Gamma(1+c)}[x(1-x)]^c,
\end{eqnarray}
with $c=0.4$. In the nonfactorizable contributions, $S_t(x)$ gives
a very small numerical effect to the amplitude~\cite{L4}.
Therefore, we drop $S_t(x)$ in $h_n$ and $h_{na}$.

The evolution factors $E^{(\prime)}_e$ and $E^{(\prime)}_a$
entering in the expressions for the matrix elements (see section 3) are
given by
\begin{eqnarray}
E_e(t)&=&\alpha_s(t) \exp[-S_B(t)-S_3(t)],
 \ \ \ \
 E'_e(t)=\alpha_s(t)
 \exp[-S_B(t)-S_2(t)-S_3(t)]|_{b_1=b_3},\\
E_a(t)&=&\alpha_s(t)
 \exp[-S_2(t)-S_3(t)],\
 \ \ \
E'_a(t)=\alpha_s(t) \exp[-S_B(t)-S_2(t)-S_3(t)]|_{b_2=b_3},
\end{eqnarray}
in which the Sudakov exponents are defined as
\begin{eqnarray}
S_B(t)&=&s\left(x_1\frac{M_{B_s}}{\sqrt
2},b_1\right)+\frac{5}{3}\int^t_{1/b_1}\frac{d\bar \mu}{\bar
\mu}\gamma_q(\alpha_s(\bar \mu)),\\
S_2(t)&=&s\left(x_2\frac{M_{B_s}}{\sqrt
2},b_2\right)+s\left((1-x_2)\frac{M_{B_s}}{\sqrt
2},b_2\right)+2\int^t_{1/b_2}\frac{d\bar \mu}{\bar
\mu}\gamma_q(\alpha_s(\bar \mu)),
\end{eqnarray}
 with the quark
anomalous dimension $\gamma_q=-\alpha_s/\pi$. Replacing the
kinematic variables of $M_2$ to $M_3$ in $S_2$, we can get the
expression for $S_3$. The explicit form for the  function
$s(Q,b)$ is:
\begin{eqnarray}
s(Q,b)&=&~~\frac{A^{(1)}}{2\beta_{1}}\hat{q}\ln\left(\frac{\hat{q}}
{\hat{b}}\right)-
\frac{A^{(1)}}{2\beta_{1}}\left(\hat{q}-\hat{b}\right)+
\frac{A^{(2)}}{4\beta_{1}^{2}}\left(\frac{\hat{q}}{\hat{b}}-1\right)
-\left[\frac{A^{(2)}}{4\beta_{1}^{2}}-\frac{A^{(1)}}{4\beta_{1}}
\ln\left(\frac{e^{2\gamma_E-1}}{2}\right)\right]
\ln\left(\frac{\hat{q}}{\hat{b}}\right)
\nonumber \\
&&+\frac{A^{(1)}\beta_{2}}{4\beta_{1}^{3}}\hat{q}\left[
\frac{\ln(2\hat{q})+1}{\hat{q}}-\frac{\ln(2\hat{b})+1}{\hat{b}}\right]
+\frac{A^{(1)}\beta_{2}}{8\beta_{1}^{3}}\left[
\ln^{2}(2\hat{q})-\ln^{2}(2\hat{b})\right],
\end{eqnarray} where the variables are defined by
\begin{eqnarray}
\hat q\equiv \mbox{ln}[Q/(\sqrt 2\Lambda)],~~~ \hat b\equiv
\mbox{ln}[1/(b\Lambda)], \end{eqnarray} and the coefficients
$A^{(i)}$ and $\beta_i$ are \begin{eqnarray}
\beta_1=\frac{33-2n_f}{12},~~\beta_2=\frac{153-19n_f}{24},\nonumber\\
A^{(1)}=\frac{4}{3},~~A^{(2)}=\frac{67}{9}
-\frac{\pi^2}{3}-\frac{10}{27}n_f+\frac{8}{3}\beta_1\mbox{ln}(\frac{1}{2}e^{\gamma_E}),
\end{eqnarray}
$n_f$ is the number of the quark flavors and $\gamma_E$ is the
Euler constant. We will use the one-loop running coupling
constant, i.e. we pick up the four terms in the first line of the
expression for the function $s(Q,b)$.



\end{spacing}

\begin{thebibliography}{}
\bibitem{cab}N. Cabibbo, Phys. Rev. Lett. {\bf10}, 531 (1963).
\bibitem{kob}M. Kobayashi and T. Maskawa, Prog.
Theor. Phys. {\bf49}, 652 (1973).
\bibitem{J-R.A} R. Aaij et al. (LHCb Collaboration), Phys. Rev. Lett. {\bf111}, 101801 (2013);
Phys. Rev. Lett. {\bf112}, 011801 (2014).
\bibitem{HB1997}H.B. O'Connell, B.C.Pearce,
A.W. Thomas, and A.G. Williams, Prog. Part. Nucl. Phys. {\bf 39}, 201 (1997);
H.B. O'Connell, Aust. J. Phys. {\bf 50}, 255 (1997).
\bibitem{Ryoji1996}Ryoji Enomoto, Masaharu Tanabashi, Phys. Lett. {\bf B386}, 413 (1996).
\bibitem{Rebecca2017}Rebecca Klein, Thomas Mannel, Javier Virto, and K. Keri Vos, JHEP {\bf 1710}, 117 (2017).
\bibitem{Yali2017}Ya Li, Ai-Jun, Wen-Fei Wang, and Zhen-Jun Xiao, Phys. Rev. {\bf D95}, 056008 (2017).

\bibitem{AC}A. Ali and C. Greub, Phys. Rev. {\bf D57}, 2996 (1998); G. Kramer, W. F. Palmer, and H. Simma, Nucl. Phys. {\bf B428}, 77 (1994); Z. Phys. {\bf C66}, 429 (1995).



\bibitem{AG}A. Ali, G. Kramer, and C.-D. Lv, Phys. Rev. {\bf D58}, 094009 (1998); {\bf59}, 014005 (1998); Y. H. Chen, H. Y. Cheng, B. Tseng, and K. C. Yang, Phys. Rev. {\bf D60}, 094014 (1999).
\bibitem{YA}Y. Y. Keum, H.-n. Li, and A. I. Sanda, Phys. Rev. Lett. {\bf B504}, 6 (2001); Phys. Rev. {\bf D63}, 054008 (2001).
\bibitem{LKM}C.-D. Lv, K. Ukai, and M.-Z. Yang, Phys. Rev. D{\bf63}, 074009 (2001).
\bibitem{GW}G. Buchalla, A. J. Buras, M. E. Lautenbacher, Rev. Mod. Phy. {\bf68}, 1125 (1996).



\bibitem{gar}S. Gardner, H.B. O'Connell, and A.W. Thomas, Phys.
Rev. Lett. {\bf 80}, 1834 (1998).
\bibitem{guo1}X.-H. Guo and A.W. Thomas, Phys. Rev. {\bf D58}, 096013 (1998).

\bibitem{guo2}X.-H. Guo, O. Leitner, and A.W.Thomas, Phys. Rev. {\bf D63}, 056012 (2001).
\bibitem{guo11}X.-H. Guo and A.W. Thomas, Phys. Rev. {\bf D61}, 116009 (2000).
\bibitem{lei}O. Leitner, X.-H. Guo, and A.W. Thomas, Eur. Phys. J.
{\bf C31}, 215 (2003).
\bibitem{gang1}X.-H. Guo, Gang L$\ddot{u}$ and Z.-H. Zhang, Eur. Phys. J.
{\bf C58}, 223 (2008).
\bibitem{gang2}Gang L$\ddot{u}$, Bao-He Yuan, Ke-Wei Wei, Phys. Rev. {\bf D83}, 014002 (2011).
\bibitem{gang3}Gang L$\ddot{u}$, Zhen-Hua Zhang, Xiu-Ying Liu and Li-Ying Zhang,
Int. J. Mod. Phys. {\bf A26}, 2899 (2011).





\bibitem{CeK}C.E. Wolfe, K.Maltman, Phys. Rev. {\bf D80}, 114024 (2009).
\bibitem{CEK}C.E. Wolfe, K.Maltman, Phys. Rev. {\bf D83}, 077301 (2011).
\bibitem{gang4}Gang L$\ddot{u}$, Ye Lu, Sheng-Tao Li and Yu-Ting Wang,
Eur. Phys. J. {\bf C77}, 518 (2017).
\bibitem{wol}L. Wolfenstein, Phys. Rev. Lett. {\bf 51}, 1945 (1983); Phys. Rev. Lett. {\bf13}, 562 (1964).
\bibitem{18} V. M. Braun and A. Lenz,  Phys. Rev. {\bf D70}, 074020 (2004); P. Ball and A. Talbot,
JHEP 0506, 063 (2005); P. Ball and R. Zwicky, Phys. Lett. {\bf B633}, 289 (2006)
; A. Khodjamirian, Th. Mannel and M. Melcher, Phys. Rev. {\bf D70}, 094002 (2004).

\bibitem{19} T. Feldmann, P. Kroll and B. Stech, Phys. Rev. {\bf D58}, 114006 (1998).

\bibitem{PDG2016}C. Patrignani et al. (Particle Data Group), Chin. Phys. {\bf C40}, 100001 (2016).

\bibitem{ADR2016}A. Bharucha, D. M. Straub, R. Zwicky, JHEP, 1608, 098 (2016).

\bibitem{XZZ2016}X. Liu, Z. J. Xiao, Z. T. Zou, Phys. Rev. {\bf D94}, 113005 (2016).


\bibitem{AMLi2007}Ahmed Ali, Gustav Kramer, Ying Li, Cai-Dian Lu, Yue-Long Shen, Wei Wang, Phys. Rev. {\bf D76}, 074018 (2007).


\bibitem{L3} H.-n. Li, Phys. Rev. {\bf D66}, 094010 (2002).

\bibitem{L4} H.-n. Li and K. Ukai, Phys. Lett. {\bf B555}, 197 (2003).

\end{thebibliography}
\end{document}